\documentclass[12pt]{iopart}
\usepackage{graphicx}
\usepackage{xcolor}

\begin{document}

\title[A network model of pulmonary fibrosis]{A network model of
  correlated growth of tissue stiffening in pulmonary fibrosis}

\author{Cl\'audio L N Oliveira$^1$, Jason H T Bates$^2$ and B\'ela
  Suki$^1$} \address{$^1$ Department of Biomedical Engineering, Boston
  University, Boston, MA 02215} \address{$^2$ Department of Medicine,
  University of Vermont, Burlington, VT 05405}
\eads{\mailto{lucas@fisica.ufc.br},
  \mailto{jason.h.bates@med.uvm.edu}, \mailto{bsuki@bu.edu}}

\begin{abstract}
  During the progression of pulmonary fibrosis, initially isolated
  regions of high stiffness form and grow in the lung tissue due to
  collagen deposition by fibroblast cells. We have previously shown
  that ongoing collagen deposition may not lead to significant
  increases in the bulk modulus of the lung until these local
  remodeled regions have become sufficiently numerous and extensive to
  percolate in a continuous path across the entire tissue [Bates {\it
    et al.} 2007 {\it Am. J. Respir. Crit.  Care Med.}  {\bf 176}
  617].  This model, however, did not include the possibility of
  spatially correlated deposition of collagen. In the present study,
  we investigate whether spatial correlations influence the bulk
  modulus in a two-dimensional elastic network model of lung tissue.
  Random collagen deposition at a single site is modeled by increasing
  the elastic constant of the spring at that site by a factor of 100.
  By contrast, correlated collagen deposition is represented by
  stiffening the springs encountered along a random walk starting from
  some initial spring, the rationale being that excess collagen
  deposition is more likely in the vicinity of an already stiff
  region. A combination of random and correlated deposition is modeled
  by performing random walks of length $N$ from randomly selected
  initial sites, the balance between the two processes being
  determined by $N$. We found that the dependence of bulk modulus,
  $B(N,c)$, on both $N$ and the fraction of stiff springs, $c$, can be
  described by a strikingly simple set of empirical equations. For
  $c<0.3$, $B(N,c)$ exhibits exponential growth from its initial value
  according to $B(N,c)\approx B_0exp(2c)[1+c^{\beta}\ln(N^{a_I})]$,
  where $\beta=0.994\pm 0.024$ and $a_I=0.54\pm 0.026$. For
  intermediate concentrations of stiffening, $0.3\le c\le 0.8$,
  another exponential rule describes the bulk modulus as
  $B(N,c)=4B_0exp[a_{II}(c-c_c)]$, where $a_{II}$ and $c_c$ are
  parameters that depend on $N$. For $c>0.8$, $B(N,c)$ is linear in
  $c$ and independent of $N$, such that
  $B(N,c)=100B_0-100a_{III}(1-c)B_0$, where $a_{III}=2.857$. For small
  concentrations, the physiologically most relevant regime, the forces
  in the network springs are distributed according to a power law.
  When $c=0.3$, the exponent of this power law increases from -4.5,
  when $N=1$, and saturates to about -2, as $N$ increases above 40.
  These results suggest that the spatial correlation of collagen
  deposition in the fibrotic lung has a strong effect on the rate of
  lung function decline and on the mechanical environment in which the
  cells responsible for remodeling find themselves.
\end{abstract}


\section{Introduction}
Pulmonary fibrosis (PF) is a debilitating disease without cure that
affects approximately 5 million people worldwide~\cite{Meltzer2008}
and leads to the death of 40,000 people every year in the US
alone~\cite{PFwebsite}. Anti-inflammatory, antifibrotic, and
immunosuppressive therapies have all been used in the treatment of PF,
but unfortunately none have been shown to improve survival or quality
of life~\cite{Raghu2004,Egan1999}. The only truly effective treatment
for PF remains lung transplantation, but donor lungs are scarce, so
being able to accurately predict the stage of the disease is critical
for deciding how best to choose recipients for donor lungs among the
many patients who could potentially benefit.

A key clinical indicator of PF severity is the increase it causes in
the elastic stiffness of the lung, which is readily measured in
patients~\cite{Martinez2005}. However, we have shown that the
progression of macroscopic lung stiffness may not always precisely
track the microscopic progression of the underlying fibrotic
disease~\cite{Bates2007,Suki2007,Suki2008,Suki2011,Cavalcante2005}, so
understanding exactly how these two processes are linked is clearly of
major importance for PF prognosis.  Also, the progression of PF at the
microscopic level influences the rate at which the local
microenvironment of the tissue changes, which in turn could influence
the rate of fibrotic remodeling by lung fibroblast cells.
Understanding the nature of this progression may thus be crucial for
the development of novel therapies.

Some cases of PF have a genetic basis, some result from known injuries
to the lung tissue, and many are idiopathic~\cite{Levitzky1995}.
Regardless of the cause, however, the evolution of PF is invariably
accompanied by a progressive stiffening of the lung tissue that
manifests clinically as a reduction in lung volume and an increase in
the work of breathing.  Microscopically, PF is inherently a remodeling
process in which, at least initially, isolated local regions of the
lung become fibrotic as collagen is deposited by the fibroblast cells.
This increases regional tissue stiffness, but the effect on overall
lung stiffness is more subtle. In particular, we have shown
previously~\cite{Bates2007,Suki2007,Suki2008,Suki2011,Cavalcante2005}
that the total amount of collagen deposited in the lung during PF may
be a poor indicator of abnormalities in global lung stiffness as
fibrosis develops. More important is whether or not the stiffened
tissue remains in isolated islands or becomes linked in continuous
paths that span the entire tissue space. When such paths arise, at the
percolation threshold~\cite{Stauffer1992,Andrade2011}, the subsequent
rate of increase in lung stiffness becomes dramatically enhanced as PF
progresses, which represents the rapid approach of end-stage disease.

\begin{figure}[t]
  \begin{center}
    \includegraphics*[width=13.0cm]{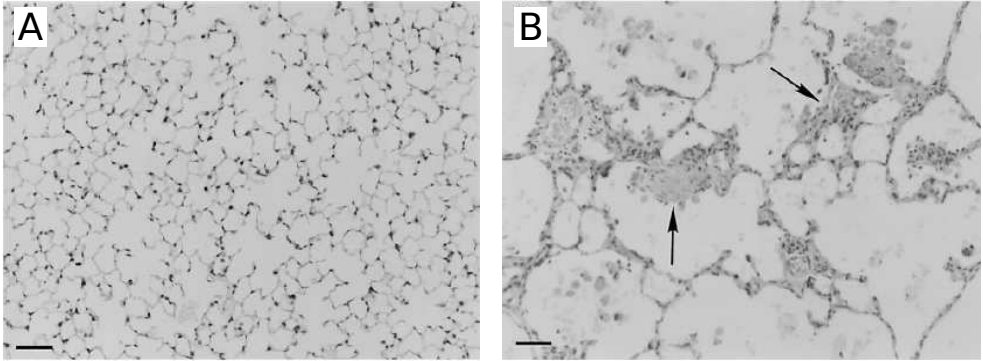}
    \caption{Microscopic images of normal (A) and fibrotic (B) tissues
      in the lung. Note in the fibrotic lung the significant
      heterogeneities in alveolar sizes (light gray areas) due to the
      formation of fibrotic lesions (dark gray clusters shown by the
      arrows). These fibrotic regions create characteristic scarring
      patterns. Scale bars are 50 $\mu m$ size. The images are
      reproduced with permission from Ref.~\cite{Hoyle1999}.
      (permission is under request)}\label{f.images}
  \end{center}
\end{figure}

The point at which the percolation threshold is reached for a given
collagen load is strongly influenced by the spatial pattern of the
collagen deposition. The characteristic scarring patterns seen in PF
indicate that this deposition does not manifest as a large number of
lesions of all different sizes, but rather as a relatively small
number of larger fibrotic patches (Figure~\ref{f.images}). This would
seem to indicate that fibrosis development in the lung is not random,
but rather occurs with a significant degree of spatial correlation.
One possible explanation for this spatial correlation is the existence
of fixed sites of fibrogenesis that spawn radial growth of collagen
deposition.  Another possibility is that once collagen is deposited at
a particular location, the local increase in stiffness that it
engenders gives rise to a pro-fibrotic mechanical environment for the
surrounding cells which themselves start to produce collagen in a
positive feedback loop. Understanding the link between total collagen
load and lung stiffness may thus provide a clue as to which of these
two possibilities is more likely to pertain in PF.

In the present study, we build on our prior work on the percolation of
random fibrotic lesions across the
lung~\cite{Bates2007,Suki2007,Suki2008,Suki2011,Cavalcante2005} to
introduce a new elastic network model of the fibrotic lung that
enables us to investigate how the evolution of lung tissue stiffness
depends on the spatial correlation of fibrotic lesions as they develop
over time. We use random placement of stiffened springs in the network
to mimic the uncorrelated appearance of new fibrotic regions, and the
stiffening of springs along a random walk~\cite{Havlin2000} from some
starting point to mimic the correlated local growth of fibrosis.
Invoking both {\it percolation} and {\it random walk} simultaneously
represents the intrinsic combination of both random and correlated
depositions of collagen. The relative contributions of these two
processes are controlled by specifying the length of the random walk.
We show here that the length of the random walk significantly affects
the spatial organization of fibrotic regions in the model, which in
turn influences the evolution of overall model stiffness.

\section{Model}
\label{model}
Our model is a two-dimensional network of hexagonally arranged linear
elastic springs, each spring representing an alveolar wall as seen in
a thin slice of lung tissue. The outer borders of the network are held
fixed such that each spring is pre-stressed to correspond to a given
lung volume.  The model is initialized to be homogeneous with all
springs having identical unit stiffness. Local collagen deposition is
modeled by stiffening individual springs by sequentially applying the
following two growth rules:
\begin{enumerate}
\item {\it Rule 1 -- Percolation}: A spring is chosen at random and
  stiffened by a factor of 100 representing the initial deposition of
  collagen;
\item {\it Rule 2 -- Random Walk}: A spring is chosen at random from
  the neighbors of the spring stiffened in Rule 1. This spring is
  stiffened by a factor of 100, followed by the stiffening of a random
  neighbor, and so on. After $N-1$ springs have been visited along
  this random walk, the new equilibrium configuration of the network
  is determined. This random walk represents the local spreading of
  fibrosis.
\end{enumerate} 
This sequence of applying {\it Rules 1} and {\it 2} followed by
finding the minimum energy configuration of the network is continued
until all springs in the network have been stiffened.

{\it Rule 1} is the same uncorrelated regular percolation that we
introduced
previously~\cite{Bates2007,Suki2007,Suki2008,Suki2011,Cavalcante2005}.
{\it Rule 2} incorporates random springs from {\it Rule 1} to seed the
initial sites of spring stiffening, and for $N=1$, the length of the
random walk is zero. For $N\gg 1$, however, most of the spring
stiffening in {\it Rule 2} occurs along random walk trajectories
starting from the initial seed locations. This gives rise to local
clusters of stiffened springs that tend to have an irregular shape.
Also, since a random walk is a Markovian process, a spring may be
visited more than once in a given random walk, resulting in cluster
sizes less than or equal to $N$ springs. Note that a spring's
stiffness is only allowed to increase once, regardless of how many
times it has been visited.

The biological rationales motivating these rules are as follows.
First, we chose to stiffen each spring by a factor of 100 because this
eventually increases the local stiffness similar to what has been
reported for fibrotic human lung tissue~\cite{Liu2011,Discher2005}.
{\it Rule 1} assumes that fibrosis occurs purely by chance, and that
its presence does not contribute to the fibrotic mechanism. {\it Rule
  2}, on the other hand, assumes that the local presence of fibrosis
has a significant influence on the occurrence of further fibrosis in
the affected region. A possible mechanism for this influence is
chemical; any mediators involved in fibrogenesis are likely to diffuse
throughout the neighboring region and thus grow a fibrotic lesion
outward from its starting point.  However, the spread of fibrosis can
also be due to mechanical mechanisms. For example, a stiff fibrotic
patch is likely to increase the stress and strain experienced by
neighboring normal tissue as lung volume cycles during breathing. If a
fibroblast cell sitting in this neighboring tissue responds to this
change in environment by depositing extra collagen as a defense
mechanism against the extra strain, then a positive feedback mechanism
will result that again could cause fibrosis to spread outward from the
site of initiation.  The value of $N$ in {\it Rule 2} essentially
controls the feedback gain.

\begin{figure}[t]
  \begin{center}
    \includegraphics*[width=13.cm]{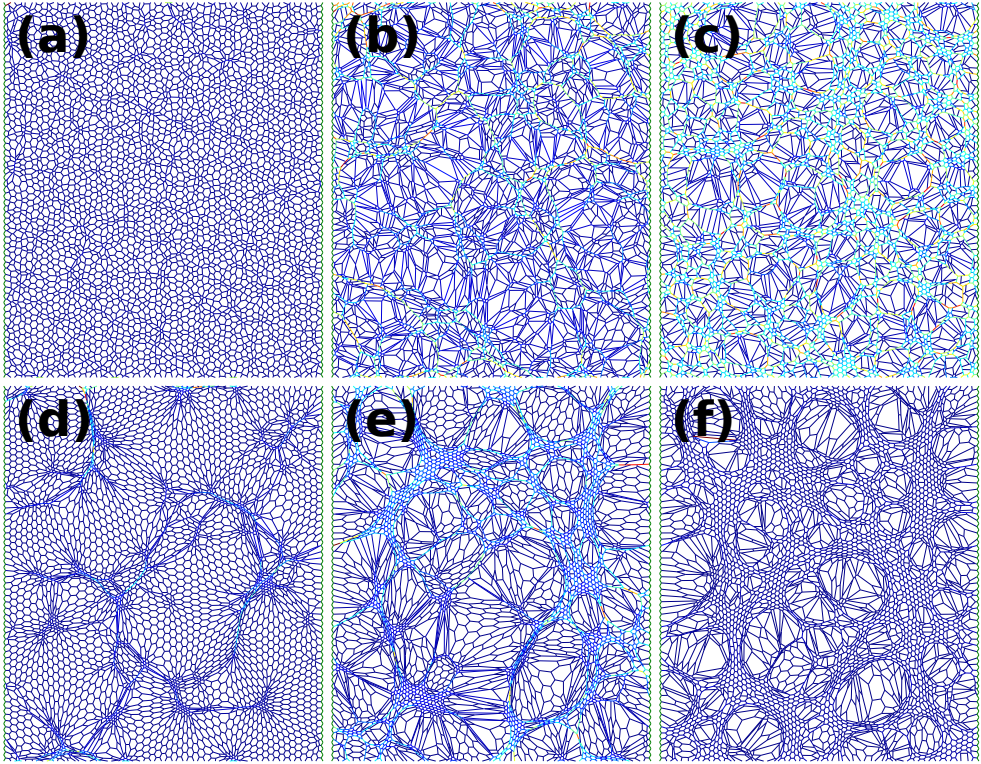}
    \caption{Snapshots of the network for $N=1$ (a, b, and c) and
      $N=100$ (d, e, and f) showing the evolution of fibrotic clusters
      with the fraction, $c$, of stiffened springs given by $c=0.3$ (a
      and d), $c=0.65$ (b and e), and $c=0.8$ (c and f).  The colors
      represent the magnitude of the spring force, varying from blue
      (low) to red (high). Scarring patterns, as those seen in
      Figure~\ref{f.images}, are better represented when random walk
      ($N>1$) is included in the model.}\label{f.MT}
  \end{center}
\end{figure}

At every step following the application of {\it Rule 1} and {\it Rule
  2}, we use a method based on {\it simulated
  annealing}~\cite{Kirkpatrick1983} to determine the configuration of
the network for which its total energy, $E$, is minimized. The
configuration of the network is defined by the positions of its nodes
(the points where the springs interconnect).  $E$ is given by the sum
of the elastic energies in all the springs in the network, i.e.
\begin{equation}
  E = \sum_i\frac{k_i\Delta\ell_i^2}{2},
\end{equation}
where $\Delta\ell_i$ and $k_i$ are the extension and stiffness,
respectively, of the $i^{th}$ spring. Starting from some initial
configuration, we determine the equilibrium (energy minimizing)
configuration in an iterative procedure as follows. At each step in
the procedure, each non-fixed node is allowed to move by a small
amount in the direction of the resultant force exerted on it by the
three springs to which it is connected. This causes $E$ to change by a
small amount $\Delta E$. If $\Delta E$ is negative (i.e. energy is
less in the new configuration) then the new configuration is accepted.
If $\Delta E$ is positive then the new configuration is accepted with
a small probability given by $p=exp(-\Delta E/T)$ where $T$ is a
control parameter. This occasional acceptance of increased energy
configurations allows the system the chance to escape from local
energy minima in search of the global minimum. We repeated this
stochastic process until $\Delta E/E$ remained below $10^{-7}$ for 20
consecutive iterations, at which point we considered the algorithm to
have converged. We performed this energy minimization procedure only
after the completion of each random walk, which assumes that any
network configurations taking place during the walk did not influence
the walk itself.

We simulated networks of $81\times 40$ nodes, as shown in
Figure~\ref{f.MT} which illustrates the influence of $N$ on the
macroscopic appearance of the fibrotic regions. Each curve in this and
the other figures presented below was obtained by first determining
the bulk modulus for 30 independent network realizations for a given
value of $N$. Linear interpolation was then applied to each
relationship to determine the value of the bulk modulus at a fixed set
of equally spaced values of $0<c<1$. Finally, the 30 interpolated
relationships were ensemble averaged. Comparison of
Figures~\ref{f.images} and \ref{f.MT} suggests that $N=100$ gives rise
to more realistic scarring patterns than does $N=1$.

\begin{figure}[t]
  \begin{center}
    \includegraphics*[width=11.cm]{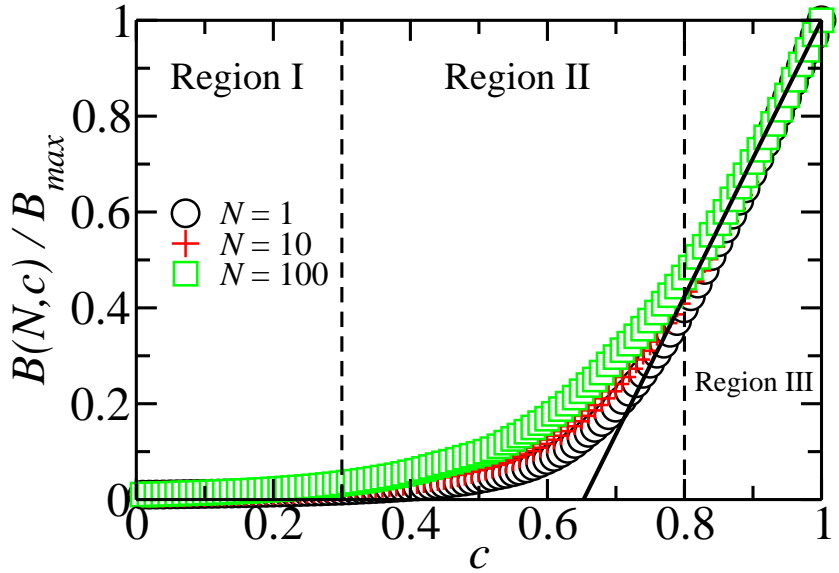}
    \caption{The bulk modulus, $B(N,c)$, as a function of $N$ versus
      the concentration of stiff springs, $c$, for $N$ = 1, 10, and
      100.  $B(N,c)$ is normalized by the maximum value of the bulk
      modulus, $B_{max}$, when all springs are stiff ($c=1$).  Three
      regimes of mechanical behavior can be distinguished according to
      the concentration: $c<0.3$ (Region I); $0.3\le c\le 0.8$ (Region
      II); and $c>0.8$ (Region III).  Each of these regions presents a
      particular growth behavior for $B(N,c)$.  The solid line is a
      straight line between the coordinates (0.65,0) and (1,1).}
    \label{f.BulkMod}
  \end{center}
\end{figure}

\section{Bulk Modulus}
\label{bulk}
In order to quantify how spatial correlations in the growth of
fibrotic regions affect the stiffness of the entire spring network, we
calculated the two-dimensional averaged bulk modulus, $B$, of the
network as a function of the fraction, $c$, of stiffened springs for
different values of $N$. The value of $c$ spans the range from 0
(completely healthy tissue) to 1 (entirely fibrotic tissue). For each
realization, the bulk modulus was computed for each value of $N$ and
$c$ by temporarily increasing the network boundaries in both
directions by 0.001\% and finding the stretched equilibrium
configuration. The bulk modulus was then taken as the ratio of the
change in stress\footnote{Stress is defined here as the sum of
  perpendicular force, per unit length, in both direction in each
  spring.} to the change in the area defined by the network boundaries
following stretch.  The bulk modulus is thus equivalent to the inverse
of specific lung compliance used in physiological studies.  Note that
$B(1,c)$ is determined using {\it Rule 1} because only the initial
seed spring is stiffened at each step in this rule, while $B(N,c)$ for
$N>1$ is determined using both {\it Rule 1} and {\it Rule 2}. We also
define $B_0$ as the minimum value of $B$ that pertains to the healthy
network containing only soft springs, while $B_{max}$ is the maximum
$B$ that occurs when all the springs have been stiffened. These two
constants, $B_0$ and $B_{max}$, apply only to fully homogeneous
networks and, therefore, are independent on $N$.  Moreover, since we
considered only Hookean springs, we have that $B_{max}=100B_0$. We
find that $B(N,c)$ segregates into three distinct regimes (see
Figure~\ref{f.BulkMod}); Region I corresponds to $c<0.3$, Region II
corresponds to $0.3\le c\le 0.8$, and Region III corresponds to
$c>0.8$. In Regions I and II, $B(N,c)$ exhibits an exponential
dependence on $c$ and also depends on $N$. By contrast, in region III,
$B(N,c)$ exhibits a linear dependence on $c$ and is independent of
$N$. We chose these regions in such a way that we could provide simple
empirical equations for each one of them. These relationships are then
demonstrated in the following subsections.

\begin{figure}[t]
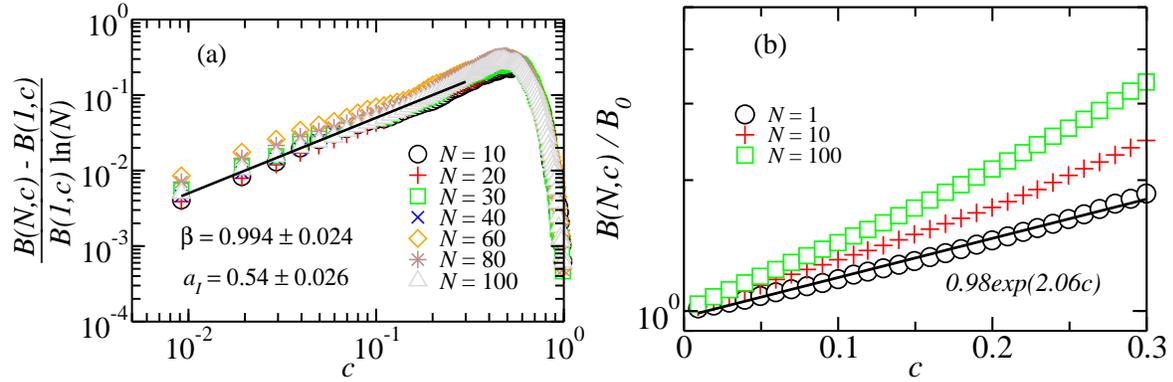

  \begin{center}
    \includegraphics*[width=7.6cm]{concxDifstress2_new.eps}
    \includegraphics*[width=7.6cm]{BulkModulus3.eps}
    \caption{A data collapse is obtained by subtracting $B(1,c)$ from
      $B(N,c)$ and divided by $B(1,c)\ln(N)$ as shown in graph (a).
      For $c<0.3$ (Region I), this collapse follows a power law (solid
      line), with some finite-size effects in the form of
      [$B(N,c)-B(1,c)]/[B(1,c)\ln(N)]=a_Ic^{\beta}$, where
      $\beta=0.994\pm 0.024$ and $a_I=0.54\pm 0.026$. In graph (b),
      $B(N,c)$ starts from $B_0$ for $c=0$ and grows with $c$ for the
      log-linear plot of the normalized $B(N,c)$ versus $c$, for $N$ =
      1, 10, and 100.  The fit (solid line) shows that $B(1,c)$ grows
      exponentially according to $B(1,c)=0.98B_0exp(2.06c)$. The graph
      also shows that a greater $N$ produces a faster rise of
      $B(N,c)$.}
    \label{f.BulkMod_R1}
  \end{center}
\end{figure}

\subsection{Region I}
For low concentrations of stiffened springs, when $c<0.3$, $B(N,c)$
starts from the initial minimum value $B_0$ at $c=0$, regardless of
$N$, and increases exponentially according to the relationship
\begin{equation}
  B(N,c) = B(1,c)[1+c^{\beta}\ln(N^{a_I})],
  \label{e.BNI}
\end{equation}
where $\beta=0.994\pm 0.024$ and $a_I=0.54\pm 0.026$. We obtained this
expression by fitting a power law to $[B(N,c)-B(1,c)]/[B(1,c)\ln(N)]$
versus $c$ as shown in Figure~\ref{f.BulkMod_R1}(a). For $N=1$, the
logarithm in Equation~\ref{e.BNI} vanishes giving $B(N,c)=B(1,c)$,
which is described by $B(1,c)=0.98B_0exp(2.06c)$ (see
Figure~\ref{f.BulkMod_R1}(b)).

Equation~\ref{e.BNI} shows that one can separate the influence of
random percolation and random walk into two independent terms. That
is, $B(N,c)=f_{p}f_{rw}$ where $f_{p}$ is the percolation term
$B(1,c)$ and $f_{rw}$ is the random walk term
$[1+c^{\beta}\ln(N^{a_I})]$.  This partitioning is possible because,
when $c$ is small and stiffened springs are sparse, the probability
that two fibrotic clusters will meet each other during growth is
small.  Interestingly, given that $a_I\approx 0.5$, the logarithmic
term grows roughly with $\sqrt N$, the average diameter of the area
covered by the random walk.

If we replace $B(1,c)$ in Equation~\ref{e.BNI} with its expression
derived from Figure~\ref{f.BulkMod_R1}(b), which shows that
$B(1,c)\approx B_0exp(2c)$, and replace $\beta$ and $a_I$ by their
approximated values 1 and 0.5, respectively, then Equation~\ref{e.BNI}
becomes
\begin{equation}
  B(N,c)\approx B_0exp(2c)[1+c\ln(\sqrt N)].
\end{equation}
Notice that for $c=0$ one obtains $B(N,c)=B_0$, as expected.

\begin{figure}[t]
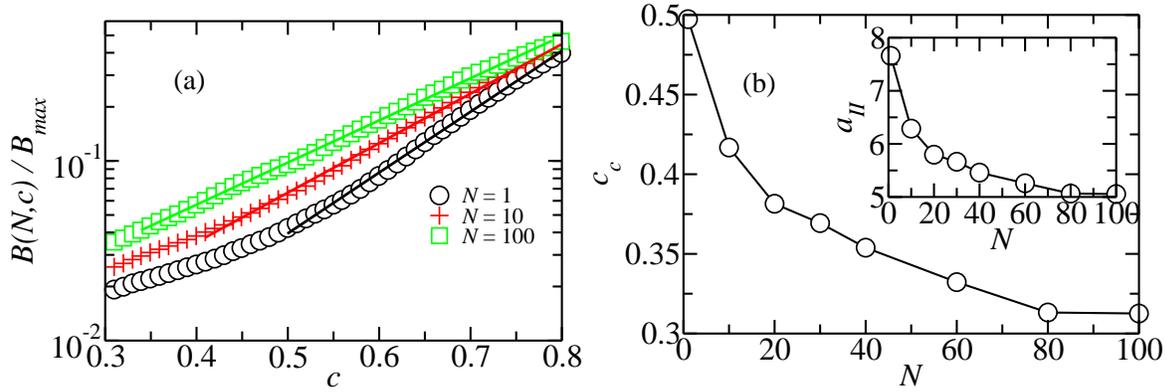

  \begin{center}
    \includegraphics*[width=7.6cm]{BulkModulus2.eps}
    \includegraphics*[width=7.6cm]{ccxN.eps}
    \caption{For intermediate concentrations (Region II), graph (a)
      shows that the bulk modulus follows an exponential with
      $N$-dependent parameters,
      $B(N,c)=0.04B_{max}exp[a_{II}(c-c_c)]$, where $a_{II}$ and $c_c$
      decrease with $N$ as shown in graph (b). The parameter $c_c$ is
      the value of the concentration where this fit starts. For $N\gg
      1$, $c_c$ approaches 0.3 and the fit spans the entire Region II
      ($0.3\le c\le 0.8$).}
    \label{f.BulkMod_R2}
  \end{center}
\end{figure}

\subsection{Region II}
Above $c=0.3$, $B(N,c)$ changes its dependence on $c$ to a different
exponential relationship. The new behavior starts at a value of
$c=c_c$ that varies inversely with $N$ (see solid lines in
Figure~\ref{f.BulkMod_R2}(a)), such that for $N=1$, 10, and 100, these
new exponential regimes begin at $c_c=0.5$, 0.41 and 0.34,
respectively, and these relations hold up to $c=0.8$. For $c>c_c$
these exponential relations can be expressed as
\begin{equation}
  B(N,c) = 0.04B_{max}exp[a_{II}(c-c_c)],
  \label{e.BNII}
\end{equation}
where $a_{II}$ and $c_c$ decrease monotonically with $N$ as shown in
Figure~\ref{f.BulkMod_R2}(b). The multiplicative constant 0.04 in
Equation~\ref{e.BNII} arises because $c_c$ occurs approximately when
$B(N,c)=0.04B_{max}$, regardless of $N$. Note that
Equation~\ref{e.BNII} cannot be partitioned into two independent
terms, such as $f_{p}$ and $f_{rw}$ in the case of Region I. The
reason is that in Region II, the fibrosis is extensive enough that
fibrotic clusters have a high likelihood of intersecting and this
produces a nonlinear coupling between percolation and random walk in
the growth equation.

\subsection{Region III}
When $c$ exceeds 0.8, $B(N,c)$ converges toward a single relationship
with $c$ for all values of $N$ (Figure~\ref{f.BulkMod}). This
relationship is
\begin{equation}
  B(N,c) = B_{max}-a_{III}(1-c)B_{max},
  \label{e.BNIII}
\end{equation}
where $a_{III}=\frac{1}{1-0.65}=2.857$. Intriguingly, $c=0.65$ is
approximately the location of the percolation threshold for a 2D
hexagonal networks~\cite{Sykes1964}. The behavior of $B(N,c)$ in
Region III may have less clinical relevance than that in Regions I and
II because patients might not be able to survive into Region III.
Nevertheless, our goal here is to understand the behavior of $B(N,c)$
over its entire operating range.

\section{Distribution of spring forces in the network}
\label{forces}
\begin{figure}[t]
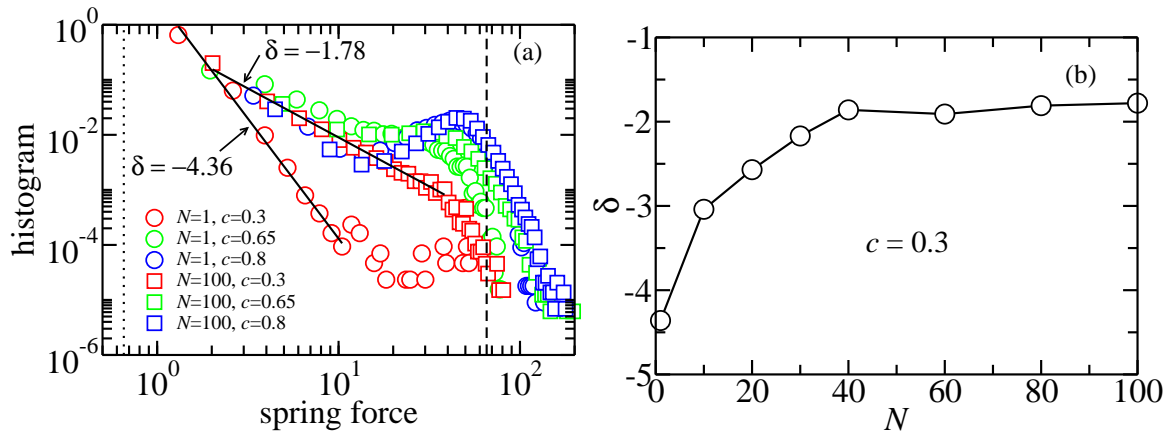

  \begin{center}
    \includegraphics*[width=7.6cm]{HistForces.eps}
    \includegraphics*[width=7.6cm]{HistForces_PLexponentxN.eps}
    \caption{The distribution of spring forces, averaged over 30
      realizations, is shown in graph (a). For the same set of
      parameters used in Figure~\ref{f.MT}, i.e., $N=1$ (circles) and
      $N=100$ (squares), and for $c=0.3$ (red), $c=0.65$ (green), and
      $c=0.8$ (blue). The dotted and dashed lines show the value of
      the spring forces in the homogeneous networks with $c=0$ and
      $c=1$, respectively. In graph (b), we show how the power law
      exponent, $\delta$, for $c=0.3$, changes with $N$. For $N>40$,
      $\delta$ fluctuates around -2.}
    \label{f.forces}
  \end{center}
\end{figure}

Next, we analyze the distribution of spring forces within the network.
Figure~\ref{f.forces}(a) shows histograms of force for $N=1$ and 100,
and for $c=0.3$, 0.65 and 0.8. The figure also indicates with a dotted
and dashed vertical line the single value of spring force in the
homogeneous networks when $c=0$ and $c=1$, respectively.

When $c=0.3$, the force distribution follows a power law with an
exponent $\delta$ that increases from -4.36 at $N=1$ toward a plateau
of about -2 for $N>40$, as shown in Figure~\ref{f.forces}(b). Hence,
for large $N$ the force distribution develops an extended tail that
leads to some regions of the network bearing more force than occurs
even when the network is fully fibrotic (i.e., $c=1$). These high
force regions correspond to single stiff springs that connect between
clusters of fibrotic regions, similarly to the red bonds in regular
percolation~\cite{Stauffer1992}.

As $c$ increases, however, $N$ loses importance in determining how the
force is distributed, as shown in Figure~\ref{f.forces}(a), and as
also occurred with $B(N,c)$ (Region III of Figure~\ref{f.BulkMod}).
Additionally, for $c\ge 0.65$, a Gaussian-like peak emerges in the
force distributions to the right of the power law decay, as shown in
Figure~\ref{f.forces}(a). Thus, the spring force distribution can be
seen to progress through a series of stages as fibrosis develops in
the network. Beginning with the single spring force of the healthy
network, Region I is accompanied by a progressively flattening power
law force distribution.  In Region II the power law distribution
starts to give way to a broad rightward peak. Eventually, in Region
III, this peak narrows to converge on the single spring force of the
fully fibrotic network.

\section{Relevance and limitations of the model}
\label{limitations}
The model of PF we have examined here is obviously based on an
extremely simple representation of real lung tissue, so we must view
our results accordingly. What these simplifications probably mean is
that although the some of the general aspects of our results may
pertain to reality, the details may not. Thus, for example, it is
reasonable to think that the nonlinear relationship we found between
the stiffening of springs and changes in $B(N,c)$
(Figure~\ref{f.BulkMod}) is likely to be mirrored in a nonlinear link
between the development of PF and clinical symptoms. Similarly, the
finding of three distinct regimes in mechanical behavior (our Regions
I, II and III) may also find analogs in reality because these regions
are defined by whether or not developing clusters of fibrosis start to
encroach on the same regions of tissue, a rather general phenomenon.
We would thus expect that at low fibrotic densities we would still
find in an actual lung that the effects of seeding density and cluster
size are separable in the manner of Equation~\ref{e.BNI}. On the other
hand, precisely where Regions I, II and III occur relative to the
progression of PF, or exactly what their respective force
distributions look like, are details that probably differ between our
model and a real fibrotic lung.

A better representation of reality could be achieved by certain
modifications to the model. For example, our network model is
2-dimensional whereas lung tissue is a 3-dimensional system in which
alveolar walls are more properly considered as elastic membranes
rather than line elements. Extending our analysis to 3-dimensions is
possible, in principle, and might prove useful for future
investigations as high-resolution 3-dimensional images of PF
progression become available. Indeed, we have previously used a
3-dimensional network model to investigate how the alveolar
architecture structure is destroyed in emphysema~\cite{Hari2011}.

Another simplification in our model is the assumption that alveolar
wall elements act as Hookean springs. This assumption of purely
elastic behavior is a simplification given that real tissue is
viscoelastic, although the long time scales over which PF develops
probably mean that the elastic assumption is adequate. The assumption
that the elasticity is linear is more questionable, given that lung
tissue exhibits marked strain stiffening. Our model thus represents a
low strain view ``zeroth''-order approach of lung stiffness. Extending
the model to include nonlinear stress-strain behavior in the spring
elements would probably not affect our overall conclusions, but would
certainly affect the details of predicted force
distributions~\cite{Ito2006}.

Finally, we simplified the process of fibrogenesis by increasing
spring constants by a single factor of 100, thereby creating a binary
stiffness map across the network, whereas it has been shown
recently~\cite{Liu2011} that the distribution of local stiffness in
the fibrotic lung is wide. While likely affecting the predicted
distributions of forces within the network, this assumption was made
in the interest of creating a simple model capable of yielding
appropriate qualitative behavior that can be readily interpreted in
terms of the interactions among its constituents.

\section{Conclusions}
\label{conclusions}
In this study, we used a simple spring network model of the lung
parenchyma to investigate how lung stiffness grows as a function of
the degree to which the deposition of collagen is spatially correlated
during the progression of PF. We simulated the progression of fibrosis
in terms of two simple growth rules, percolation and random walk, with
the degree of spatial correlation of collagen deposition being
determined by the length of the random walk. Surprisingly, we found
that the bulk stiffness of the elastic network as a function of the
amount of collagen deposition could be described by a set of three
simple empirical equations (Equations~\ref{e.BNI}, \ref{e.BNII} and
\ref{e.BNIII}), demonstrating an underlying simplicity to the way in
which percolation and random walk affect overall system behavior.

With respect to the relation between percolation and random walk, we
note the following.  Creating clusters around randomly selected points
should not change the universality class of critical properties,
since, for infinite systems, the size of such clusters can always be
reduced to single points. However, we aimed at investigating the bulk
modulus behavior when stiff clusters emerge and grow until they fill
the entire network, rather than focusing only on the critical behavior
at the percolation threshold. Moreover, bulk modulus, far away from
the criticality, depends not only on the concentration of stiff
springs but is also highly dependent on how clusters of stiff springs
spread through the network. Therefore, our results cannot be predicted
by finite-size scaling analysis alone.

Finally, we speculate that the random walk rule may represent the
homeostatic response of collagen producing cells to changes in local
stiffness as a result of the stiffening of nearby regions of tissue.
We also found that when $c$ is small, longer random walks lead to
greater heterogeneity in force distribution throughout the network
characterized by power laws. Small values of $c$ are probably the most
physiologically relevant because they represent the early to moderates
stage of fibrotic disease where patients might be expected to survive.
These results thus suggest the possibility that fibrotic remodeling in
the lung develops in part as the result of positive feedback between
collagen deposition and its consequent effects on the local mechanical
micro environment in the lung.

\ack

Supported by NIH HL-098976. Oliveira thanks the Brazilian agency CNPq
for financial support.

\end{document}